\documentclass[12pt]{article}
\def\bVec#1{\mbox{\boldmath $ #1 $}}

\usepackage[dvips]{graphicx}
\usepackage{times}
\usepackage{amsmath}
\usepackage{booktabs}
\begin{document}
\begin{center}
\begin{Large}
Dispersion relation in chiral media : \\Credibility of Drude-Born-Fedorov equations \\[5mm]
\end{Large}
Kikuo Cho \\ 
Institute of Laser Engineering, Osaka University, Suita 565-0871, Osaka Japan \\[10mm]
\end{center}

\begin{flushleft}
Abstract 
\end{flushleft}
Disperion relation of EM field in a chiral medium is discussed from the viewpoint of 
constitutive equations to be used as a partner of Maxwell equations. The popular form 
of Drude-Born-Fedorov (DBF) constitutive equations is criticized via a comparison with the 
first-principles macroscopic constitutive equations. The two sets of equations show a decisive 
difference in the dispersion curve in the resonant region of chiral, left-handed 
character, in the form of presence or absence of linear crossing at k=0. DBF equations 
could be used at most only as a phenomenology in off-resonant region, while the 
first-principles ones can be used for both phenomenological and microscopic analyses.

\section{Introduction}

Symmetry plays an important role in the electromagnetic (EM) response of matter. It is 
revealed in the form of susceptibilities relating electric and magnetic polarization 
($\bVec{P}$ and $\bVec{M}$) with source EM field.  In high symmetry case, $\bVec{P}$ 
and $\bVec{M}$ consist of (the superpositions of) independent groups of excitations
belonging to different irreducible representations of the symmetry group in consideration.  
This allows us to treat electric and magnetic properties of matter independently.  
When a medium lacks in certain mirror symmetry, i.e., the case of chiral symmetry, 
however, some (or all the) components of $\bVec{P}$ and $\bVec{M}$ cannot be  
distinguished, so that they can be induced by both electric and magnetic source fields.  
In addition, there is also a mixing between electric dipole (E1) and electric quadrupole 
(E2) transitions. 

The study of chiral symmetry in the EM response of matter has a long history 
(Introduction of \cite{Lakhtakia}).  Chiral substances have been considered as unconventional 
materials for a long time, but now it is regarded as an important source of new materials 
and states, providing hot topics in the studies of metamaterials \cite{Meta-M}, 
multiferroics \cite{multiF}, and superconductivity \cite{super-c}.

In spite of its long history, theoretical description of chirality does not 
seem to be standardized. In the 
documents of IUPAP and IUPAC dealing with the standard definitions of physical 
and chemical quantities \cite{IUPAP, IUPAC}, there is no mentioning about the 
chiral susceptibilities.  Correspondingly, there are two or more different forms 
of phenomenological constitutive equations in use for macroscopic response.  
Though the effect of chiral symmetry is expected also in microscopic responses,  
its first-principles thoery has been made only very recently \cite{Cho}.  
From the viewpoit that all the different forms of EM response theories should 
belong to a single hierarchy with logical ranking, one should be able to choose 
the most appropriate form of the constitutive equations for the macroscopic 
chiral response on the basis of the microscopic theory. 

A typical effect of chirality is the difference in the phase velocity of EM waves  
with right- and left-circular polarizations, which appears in the off-resonant 
region of susceptibilities.  However, this is not the only aspect of our interest 
in discussing chirality.  In fact, the dispersion curves in the resonant region of 
susceptibility show a remarkable behavior, by which we can select the correct 
constitutive equations.  \\

Macroscopic EM response of matter is usually calculated by the combination 
of Maxwell and constitutive equations.  The standard form of the latter is 
\begin{equation}
\label{eqn:DEBH}
 \bVec{D} = \epsilon \bVec{E} \ , \ \ \ \bVec{B} = \mu \bVec{H}
\end{equation}  
with the dielectric constant (permittivity) $\epsilon$ and permeability $\mu$.  
However, if the medium in consideration has chiral symmetry, these  constitutive 
equations need to be generalized.  A popular form of such an extention is 
\begin{eqnarray}
\label{eqn:DBFa}
  \bVec{D} &=& \epsilon (\bVec{E} + \beta \nabla \times \bVec{E} ) \ , \\
\label{eqn:DBFb}
  \bVec{B} &=& \mu (\bVec{H} + \beta \nabla \times \bVec{H} ) \ ,
\end{eqnarray}
which is called Drude-Born-Fedorov equations (DBF eqs) \cite{DBF, Band}. 
The parameter (chiral admittance) $\beta$ describes the chirality of the medium.
This is a phnomenology for uniform and isotropic media.

However, this is not the only way of generalization.  From  
the viewpoint that the fundamental variables of EM field are $\bVec{E}$ and 
$\bVec{B}$, both electric and magnetic polarizations $\bVec{P}$ and $\bVec{M}$ 
should consist both of the $\{\bVec{E}$ and $\bVec{B}\}$-induced components, 
so that the definition $\bVec{D} = \bVec{E} + 4\pi \bVec{P}$,  
$\bVec{H} = \bVec{B} - 4\pi \bVec{M}$ leads to the extension 
\begin{eqnarray}
\label{eqn:ChiCa}
  \bVec{D} &=& \hat{\epsilon} \bVec{E} + i\xi \bVec{B}  \ , \\
\label{eqn:ChiCb}
  \bVec{H} &=&  (1/\hat{\mu}) \bVec{B} + i\eta \bVec{E} \ ,
\end{eqnarray}
where the terms with $\xi$ and $\eta$ take care of the chirality.  For later 
convenience, let us call them chiral constitutive equations (ChC eqs).  Though they 
are a result of phenomenological consideration on the one hand, a first-principles 
calculation of macroscopic constitutive equations can be put also in this form 
on the other hand \cite{Cho}. 

As to the difference or similarity of DBF and ChC eqs, there is a controversy.
There have been arguments in the metamaterials community that DBF and ChC eqs 
are essentially same \cite{Engheta}, and also it is argued 
that the former can be derived from the latter by assuming the uniformity and 
isotropy of matter \cite{Lakhtakia}(Sec.4.4).  But there is also other group of people 
preferring ChC to DBF eqs.  The purpose of this article is to show that there 
is a clear difference between the two, and that ChC eqs should be preferred. 
In view of the fact that DBF eqs are frequently used in metamaterials studies  
and also in recent textbook of standard electromagnetism \cite{Band}, 
it will be important to clarify the difference between DBF and ChC eqs.  \\

We first note the relation between the parameters of DBF and ChC eqs. By means 
of the relation  
\begin{equation}
\label{eqn:AmpFar}
  \nabla \times \bVec{E} = (i\omega/c) \bVec{B} \ , \ \ \ 
  \nabla \times \bVec{H} = (-i\omega/c) \bVec{D} \ ,
\end{equation}
DBF eqs can be rewritten as 
\begin{eqnarray}
\label{eqn:DBFc}
  \bVec{D} &=& \epsilon \bVec{E} + (i\omega/c) \epsilon \beta \bVec{B} \ , \\
\label{eqn:DBFd}
  \bVec{H} &=& (i\omega/c)\epsilon \beta \bVec{E} 
                      + (1/\mu)[1 - (\omega \beta/c)^2\epsilon \mu]\bVec{B} \ .
\end{eqnarray}
If DBF and ChC eqs are equivalent, the DBF parameters can be written 
in terms of the ChC parameters by comparing (\ref{eqn:DBFc}) and 
(\ref{eqn:DBFd}) with (\ref{eqn:ChiCa}) and (\ref{eqn:ChiCb}) as 
\begin{equation} 
\label{eqn:ChC-DBF}
 \hat{\epsilon} = \epsilon \ , \ \ \ \xi = \eta = (\omega/c)\epsilon\beta\ , 
 \ \ \ (1/\hat{\mu}) = (1/\mu) - (\omega\beta/c)^2 \epsilon \ . 
\end{equation}
This relation will be shown later to lead to contradiction, which disproves 
the equivalence of DBF and ChC eqs.  \\

The first-principles derivation of micro- and macroscopic constitutive equations is 
done in the following way \cite{Cho}. We assume 
a general form of non-relativistic Hamiltonian (including relativistic correction 
terms, such as spin-orbit interaction and spin Zeeman term, etc.) for a many particle 
system in an EM field, and calculate the microscopic current density induced by 
the EM field, which is in general given as a functional of the transverse (T) part 
of vector potential $\bVec{A}^{(\rm T)}$ and the longitudinal (L) external electric 
field $\bVec{E}_{\rm ext}^{(L)}$. The integral kernel of 
the functional is the microscopic susceptibility of a separable form with respect to 
position coordinates. When the relevant quantum mechanical states have spatial 
extension much less than the wavelength of the EM field, we may apply long wavelength 
approximation to the microscopic current density, which leads to the macroscopic 
constitutive equations to be used for macroscopic Maxwell eqs.  In this macroscopic scheme, 
we need only a single $3\times 3$ tensor to relate induced current density and source 
EM field, covering all the electric, magnetic and chiral polarizations of matter. 
This macroscopic constitutive equation is given in the form \cite{Cho}
\begin{equation}
\label{eqn:const-fp}
 \bVec{J}(\bVec{k}, \omega) = \chi_{\rm em}(\bVec{k}, \omega)\ \{\bVec{A}^{(\rm T)}(\bVec{k}, \omega) 
                              - (ic/\omega) \bVec{E}_{\rm ext}^{(L)}(\bVec{k}, \omega) \} \ .
\end{equation}
The internal L field does not appear in the source field, since it is taken into account  
as the Coulomb potential in the matter Hamiltonian. The susceptibility $\chi_{\rm em}$ 
is written in terms of the quantum mechanical transition energies and the lower moments 
of the corresponding transition matrix elements of current density operator. 

Using the identity $\bVec{J} = -i\omega \bVec{P} + i c \bVec{k} \times \bVec{M}$ in 
Fourier representation, we can rigorously rewrite the constitutive equation into the form 
\begin{equation}
\label{eqn:const-PM}
 \bVec{P} = \chi_{\rm eE} \bVec{E} +\chi_{\rm eB} \bVec{B} \ , \ \ \ 
 \bVec{M} = \chi_{\rm mE} \bVec{E} +\chi_{\rm mB} \bVec{B}
\end{equation}
The four susceptibilities  $\chi_{\rm eE}$, $\chi_{\rm eB}$, $\chi_{\rm mE}$, $\chi_{\rm mB}$ 
are again written in terms of the quantum mechanical transition energies and lower transition 
moments of electric dipole (E1), electric quadrupole (E2), and magnetic dipole (M1) characters. 
Details are given in sec.3.1 of \cite{Cho}. The lowest order terms of them are
\begin{eqnarray}
 \chi_{\rm eE} &=& \frac{1}{\omega^2 V} \sum_{\nu} \big[ 
     \bar{g}_{\nu} \bar{\bVec{J}}_{0\nu} \bar{\bVec{J}}_{\nu 0} 
   + \bar{h}_{\nu} \bar{\bVec{J}}_{\nu 0} \bar{\bVec{J}}_{0\nu} \big] 
                                                         \ ,  \nonumber \\                              
 \chi_{\rm mB} &=& \frac{1}{V} \sum_{\nu} \big[ 
     \bar{g}_{\nu} \bar{\bVec{M}}_{0\nu} \bar{\bVec{M}}_{\nu 0} 
   + \bar{h}_{\nu} \bar{\bVec{M}}_{\nu 0} \bar{\bVec{M}}_{0\nu} \big]                        
                                                       \ ,       \nonumber \\
 \chi_{\rm eB} &=& \frac{i}{\omega V} \sum_{\nu} \big[ 
     \bar{g}_{\nu} \bar{\bVec{J}}_{0\nu} \bar{\bVec{M}}_{\nu 0} 
   + \bar{h}_{\nu} \bar{\bVec{J}}_{\nu 0} \bar{\bVec{M}}_{0\nu} \big]                        
                                                       \ ,       \nonumber \\
 \chi_{\rm mE} &=& \frac{-i}{\omega V} \sum_{\nu} \big[ 
     \bar{g}_{\nu} \bar{\bVec{M}}_{0\nu} \bar{\bVec{J}}_{\nu 0}
   + \bar{h}_{\nu} \bar{\bVec{M}}_{\nu 0} \bar{\bVec{J}}_{0\nu} \big] 
                                                       \ ,  \\           
   \bar{g}_{\nu} &=& \frac{1}{E_{\nu 0} - \hbar \omega - i 0^+} - \frac{1}{E_{\nu 0}} \ , \ \ \ 
   \bar{h}_{\nu} = \frac{1}{E_{\nu 0} + \hbar \omega + i 0^+} - \frac{1}{E_{\nu 0}} \ ,
\end{eqnarray}                  
where $V$ is the volume of a cell for periodic boundary condition to define $\bVec{k}$, and 
$\bar{\bVec{J}}_{0\nu}$ and $\bar{\bVec{M}}_{0 \nu}$ are, respectively, the E1 and M1 
transition moments of current density and (orbital and spin) magnetization operators between 
the matter eigenstates $|0\rangle$ (ground state) and $|\nu\rangle$ with transition energy 
$E_{\nu 0}$ between them.  (E2 moments appear in the $\bar{\bVec{J}}_{\mu\nu}$ terms 
in the next higher order.)  Chiral symmetry allows the existence of the transitions 
with mixed (E1 and M1) or (E1 and E2) character, leading to the $O(k^1)$ terms in 
$\chi_{\rm em}$. 

Though there appear four susceptibilities, the single susceptibility nature is intact, 
since the rewriting of (\ref{eqn:const-fp}) into (\ref{eqn:const-PM}) is reversible.  
Combining the new form of constitutive equations with the definition of $\bVec{D}$ and $\bVec{H}$, 
we obtain  
\begin{eqnarray} 
 \bVec{D} &=& \bVec{E} + 4\pi \bVec{P} = (1 + 4\pi \chi_{\rm eE}) \bVec{E} 
                                                             + 4\pi \chi_{\rm eB} \bVec{B} \ , \\
 \bVec{H} &=& \bVec{B} - 4\pi \bVec{M} = (1 - 4\pi \chi_{\rm mB}) \bVec{B} 
                                                             - 4\pi \chi_{\rm mE} \bVec{E} \ ,
\end{eqnarray}
which is essentially equivalent to ChC eqs. The parameters of ChC eqs are given as 
\begin{equation}
 \hat{\epsilon} = 1 + 4\pi \chi_{\rm eE}  \ , \ \ 
 i \xi = 4\pi \chi_{\rm eB} \ , \ \ 
 i \eta = -4\pi \chi_{\rm mE}  \ , \ \ 
 \hat{\mu} = \frac{1}{1 - 4\pi \chi_{\rm mB}} \ .
\end{equation}
all of which are tensors, with no assumption of isotropy and homogeneity as for DBF eqs. 
It should also be noted that the poles of $\chi_{\rm mB}$, i.e., the magnetic transition 
energies, are, not the poles, but the zeros of $\hat{\mu}$. This is due to the definition 
of $\chi_{\rm mB}$,  $\bVec{M} = \chi_{\rm mB} \bVec{B}$ as required in the 
first-principles approach, in contrast to the conventional one 
$\bVec{M} = \chi_{\rm m} \bVec{H}$.  
At this stage, the ChC eqs are not a phenomenology, but a first-principles theory.  
In contrast, this kind of first-principles derivation does not exist for DBF eqs.

\section{Dispersion equation}

In order to show the difference between DBF and ChC eqs, it is 
sufficient to give a single example.  For this purpose,we compare the dispersion 
relations obtained from DBF and ChC eqs. 

\subsection{Case of DBF eqs}

If we solve DBF eqs and eq.(\ref{eqn:AmpFar}) for $\nabla \times \bVec{H}$ and 
$\nabla \times \bVec{E}$, we obtain 
\begin{eqnarray}
\label{eqn:DBF2a}
 \nabla \times \bVec{H} &=& a \bVec{H} + b \bVec{E} \ , \\
\label{eqn:DBF2b}
 \nabla \times \bVec{E} &=& d \bVec{H} + e \bVec{E} \ ,
\end{eqnarray}
where
\begin{equation}
  a = e = -\epsilon \mu \beta / \Delta \ , \ \ \  b = -i c \epsilon /\omega \Delta \ , \ \ \ 
  d = + i c \mu/ \omega \Delta \ , 
\end{equation}
and $\Delta = \epsilon \mu \beta^2 - c^2/\omega^2$. From eqs (\ref{eqn:AmpFar}), 
(\ref{eqn:DBFc}), (\ref{eqn:DBFd}) and $\nabla\cdot \bVec{B} = 0$, both $\bVec{E}$ and 
$\bVec{H}$ are transverse, so that 
\begin{equation}
\nabla \times \nabla \times \bVec{E} = k^2 \bVec{E}, \ \ \ \ \ 
\nabla \times \nabla \times \bVec{H} = k^2 \bVec{H} \ 
\end{equation}
for Fourier components.  Then, by operating $\nabla \times $ to (\ref{eqn:DBF2a}) and 
(\ref{eqn:DBF2b}), we obtain a set of homogeneous linear equations of $\bVec{H}, \bVec{E}$.
The condition for the existence of non-trivial solution gives us the dispersion equation 
\begin{equation}
  {\rm det}| k^2 {\bf 1} - {\cal A}^2 | = 0 \ ,
\end{equation}
where ${\cal A}$ is a $2\times 2$ matrix with the components $a, b, d, e$
\begin{equation}
  {\cal A} = \left[
 \begin{array}{cc}
 a & b \\ d & e 
 \end{array} \right] \ .
\end{equation}
This dispersion equation can be rewritten as 
\begin{equation}
  {\rm det}| k {\bf 1} + {\cal A} | = 0 \ \ \ {\rm or} \ \ \ {\rm det}| k {\bf 1} - {\cal A} | = 0 \ ,
\end{equation}
so that the solution is 
\begin{equation}
  k = \pm a \pm \sqrt{bd} \ ,
\end{equation}
with all the combinations of $\pm$ being allowed, which finally leads to a compact expression   
\begin{equation}
\label{eqn:disp1}
 \frac{ck}{\omega} = \pm \frac{\sqrt{\epsilon \mu}}{1 \pm (\omega\beta/c)\sqrt{\epsilon \mu}} \ .
\end{equation}
This gives two branches of dispersion curve.  In homogeneous isotropic media, 
the two modes correspond to right and left circular polarizations.  It should be noted that 
the condition for the existence of real solution is $\epsilon \mu \geq 0$.  This means that 
the left-handed medium is defined in the same way as in non-chiral medium, in contrast to 
the case of ChC eqs.

\subsection{Case of ChC eqs}

A same way of solution is possible in this case, too.  After eliminating $\bVec{D}, \bVec{B}$ 
from the ChC eqs, the solution for $\nabla \times \bVec{H}, \ \nabla \times \bVec{E}$
has a same form as the one, where $a, b, d, e$ of previous section are replaced with 
the following  $ f, g, h, j$, respectively 
\begin{eqnarray}
 f  &=& -i(\omega/c) \xi\hat{\mu}  \ , \\
 g &=& -i(\omega/c) (\hat{\epsilon} + \hat{\mu} \xi \eta)  \ , \\
 h &=& -i(\omega/c) \hat{\mu}       \ , \\
 j &=& -i(\omega/c) \hat{\mu} \eta \ .
\end{eqnarray}
Further transformation of the equations of $\nabla \times \bVec{H}, \ \nabla \times \bVec{E}$
into a set of homogeneous equations of $\bVec{E}, \bVec{H}$ allows us to obtain the dispersion 
equation, as the condition for the existence of non-trivial solution,  
\begin{equation}
\label{eqn:disp2}
   \frac{ck}{\omega} = \pm \frac{1}{2} \big[\hat{\mu}(\eta - \xi)  
                         \pm \sqrt{\{\hat{\mu}(\eta - \xi)\}^2 + 4\hat{\epsilon} \hat{\mu}}\ \big] \ ,
\end{equation}
where we take all the combinations of $\pm$.  The condition for real solutions is 
\begin{equation}
\label{eqn:real-r}
\{\hat{\mu}(\eta - \xi)\}^2 + 4\hat{\epsilon} \hat{\mu} \geq 0 \ , 
\end{equation}
which is less restrictive than non-chiral case.

\section{Discussions}

First of all, we note that, for both DBF and ChC eqs, the well-known 
result of  $(ck/\omega)^2 = \epsilon \mu$ is obtained in the absence of chirality 
($\beta = 0$ and $\xi = 0, \eta = 0$).  Also both of the constitutive equations exhibit  
the typical behavior of chiral medium, i.e., the existence of the two branches with 
polarization dependent refractive indices.  In the non-resonant region, both of them 
could be used to fit experimental results via appropriate choice of parameter values. 

\subsection{Resonant region of left-handed chiral medium}

A decisive difference appears in resonant region.  An example will be the left-handed 
behavior emerging in the neighborhood of a chiral resonance with E1-M1 mixed character. 
Such a case has been treated by the first-principles theory of macroscopic constitutive 
equation \cite{Cho} (Sec.3.8.1 and Sec.4.1.1).  It shows a pair of dispersion curves 
for left and right circularly polarized modes, which have a linear crossing at $k=0$.  
It will be a test for the phenomenologies whether such a linear crossing can be 
realized or not by choosing parameter values. 
 
The dispersion equation in the first-principles macroscopic formalism is
\begin{equation}
  0 =  {\rm det} | \frac{c^2k^2}{\omega^2} {\bf 1} -  \big[ {\bf 1} + \frac{4\pi c}{\omega^2} 
             \chi_{\rm em}^{(\rm T)}(\bVec{k}, \omega) \big] |   \ , 
\end{equation}
where  $\chi_{\rm em}^{(\rm T)}(\bVec{k}, \omega)$ is the T component of susceptibility 
tensor (sec.2.5 of \cite{Cho}).  Let us choose a chiral form of susceptibility tensor  
$\chi_{\rm em}^{(\rm T)}$ as  
\begin{equation}
   {\bf 1} + \frac{4\pi c}{\omega^2}\chi_{\rm em}^{(\rm T)} = 
       (\epsilon_{\rm b} + a' + c'k^2) {\bf 1}  + \big[
  \begin{array}{cc}
    0 & ib'k  \\  -ib'k  &  0
  \end{array}
  \big] \ ,
\end{equation}
where the terms with $a', b', c'$ represents the contribution of a pole 
$\sim 1/(\omega_{0} - \omega)$ with mixed (E1, M1) character, while 
$\epsilon_{\rm b}$ is the background dielectgric constant due to all 
the other resonances.  We assume that the resonance with mixed E1 and M1 characters 
occurs in the frquency region of $\epsilon_{\rm b} < 0$, i.e., a chiral version of 
left-handed medium. The dispersion equation is 
\begin{equation}
  \big(\frac{ck}{\omega}\big)^2 = \bar{\epsilon} \bar{\mu} 
                                \pm \bar{\beta} \bar{\mu} \ \frac{ck}{\omega} \ ,
\end{equation}
and its solution is given as 
\begin{equation}
\label{eqn:disp3}
  \frac{ck}{\omega} = \pm \frac{1}{2} \big[ \pm \bar{\beta} \bar{\mu} 
                     + \sqrt{\bar{\beta}^2 \bar{\mu}^2 + 4 \bar{\epsilon} \bar{\mu}} \ \big] \ ,
\end{equation}
where we take all the combinations of $\pm$, and 
\begin{equation}
 \bar{\beta} =  \omega b'/c \ , \ \ \  \bar{\epsilon} = \epsilon_{\rm b} + a' \ \ \  
    \bar{\mu} = 1/[1 - (\omega/c)^2 c']         \ .
\end{equation}
Noting that $b'$ is a chiral parameter corresponding to $\xi, \eta$ of the ChC eqs, 
we see that this equation is the same type as eq.(\ref{eqn:disp2}), but not as 
eq.(\ref{eqn:disp1}).  An example of this dispersion relation is given in Fig.1. 
The characteristic behavior of the dispersion curves is a linear crossing at k=0 
(Fig.4.1 of \cite{Cho}).

\begin{figure}
 \begin{center}
  \includegraphics[width=10cm,clip]{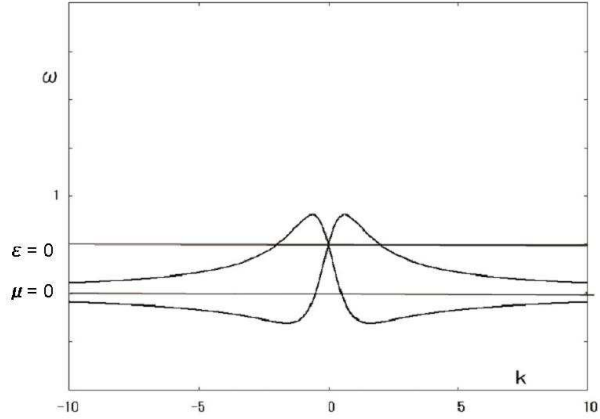}
 \end{center}
 \caption{Dispersion curves of a chiral left-handed medium for the model in the text.  
Both ordinate and abscissa are normalized by the frequency of the pole as 
$\omega/\omega_{0}$ and $ck/\omega_{0}$. Two horizontal lines show the frequncies of 
$\epsilon = 0$ and $\mu=0$.}
\label{fig:1}
\end{figure}

In order to check whether the dispersion equations (\ref{eqn:disp1}) and (\ref{eqn:disp2}) 
have this typical behavior of "linear crossing at $k=0$", we focus on the behavior of 
the dispersion equations near $k=0$. Since both of the dispersion equations are given in 
the form "$ck/\omega = F(\omega)$", we only need to examine how the function $F(\omega)$ 
approaches to zero in each case. 

The microscopic model of left-handed chiral medium given above consists of 
a matter excitation level with (E1, M1) mixed character in the frequency range of 
$\epsilon_{\rm b} < 0$.  This means that all of $\chi_{\rm eE}, 
\chi_{\rm eB},\chi_{\rm mE}, \chi_{\rm mB}$ have a common pole 
(at $\omega = \omega_{0}$) and $\chi_{\rm eE}$ is largely negative 
in the frequency range of interest to make $\epsilon_{\rm b} < 0$.  Namely, 
$\hat{\epsilon}, 1/\hat{\mu}, \xi, \eta$ of ChC eqs have a common pole at 
$\omega = \omega_{0}$ and 
$\hat{\epsilon} = \epsilon_{\rm b} + \bar{a}/(\omega_{0} - \omega)$. 
It may appear that the r.h.s. of eq.(\ref{eqn:disp2}) becoms zero for frequency 
satisfying $\hat{\mu} = 0$ or $\hat{\epsilon} = 0$.  However, as mentioned before, the zero 
of $\hat{\mu}$ corresponds to the pole of M1 transition, which in this case is common 
to the pole of $\xi$ and $\eta$. Therefore the zero of $\hat{\mu}$ is cancelled in the 
product $\hat{\mu}(\xi - \eta)$. Thus, the only remaining possibility of zero arises 
from $\hat{\epsilon} = 0$.  The $\omega$-dependence of the r.h.s. eq.(\ref{eqn:disp2})
near zero point can be found by rewriting it as 
\begin{equation}
  \pm \frac{1}{2} \frac{\hat{\epsilon} \hat{\mu}}{\hat{\mu}(\xi - \eta) \pm 
      \sqrt{\{\hat{\mu}(\xi - \eta)\}^2 + 4 \hat{\epsilon} \hat{\mu}}} \ .
\end{equation}
At the frequency satisfying $\hat{\epsilon} = 0$, i.e., $\epsilon_{\rm b} + a' = 0$, all of 
$\hat{\mu}, \xi, \eta$ remain finite, and one of the $\pm$ combinations in the denominator 
remains finite, so that the whole expression becomes zero for this combination. 
This occurs for both signs of $\pm$ in front of the whole expression. 
For negative $\epsilon_{\rm b}$ and positive numerator of the pole 
$\sim 1/(\omega_{0} - \omega)$, $\epsilon_{\rm b} + a' = 0$ occurs at 
$\omega=\omega_{\rm z} < \omega_{0}$ and 
in its neighborhood $\hat{\epsilon} \sim (\omega - \omega_{\rm z})$.  This shows that 
the r.h.s. of eq.(\ref{eqn:disp2}) behaves like $\sim (\omega - \omega_{z})$, which means 
the linear crossing of the two branches.  Note also that $\omega_{\rm z}$ lies inside the 
frequency range of eq.(\ref{eqn:real-r}).  

Now we check whether the same behavior is obtained for DBF eqs by assuming eq. 
(\ref{eqn:ChC-DBF}), from which we obtain 
\begin{equation}
 \frac{1}{\mu} = \frac{1}{\hat{\mu}}   + \frac{\xi^2}{\hat{\epsilon}} \ ,\ \ \ 
 \frac{\omega^2\beta^2}{c^2} \epsilon \mu = 1 - \frac{\mu}{\hat{\mu}} \ .
\end{equation} 
This shows that $1/\mu$ has the same pole as $1/\hat{\mu}$ at $\omega = \omega_{0}$, 
so that the factor $\mu/\hat{\mu}$ on the r.h.s. of the second equation does not  
have the pole at $\omega = \omega_{0}$ via cancellation.  Therefore, there is no 
chance for the denominator of the r.h.s. of eq.(\ref{eqn:disp1}) to diverge. Hence 
the only possibility of its becoming zero comes from the factor $\sqrt{\epsilon\mu}$ 
on the numerator.  In view of the fact that the zeros of $\epsilon, \mu$ occur at  
different $\omega$'s, e.g., at $\omega_{\rm z1}$ and $\omega_{\rm z2}$ 
($\omega_{\rm z1}$ $>$ $\omega_{\rm z2}$),  the $\omega$-dependence of 
$\sqrt{\epsilon \mu}$ should be 
$\sim \sqrt{\omega_{z1} - \omega}$ or $\sim \sqrt{\omega - \omega_{z2}}$ 
in the neigborhood of the zeros.    
Therefore, no linear crossing is possible in the DBF dispersion curves.  
The two zeros are the boundaries of the region of left-handed behavior. 

One might argue that other type of $\omega$-dependence than eq. (\ref{eqn:ChC-DBF}) 
could lead to the linear crossing behavior.  But one cannot freely give the 
$\omega$-dependence even as a phenomenology.  Linear susceptibilities should be 
a sum of single pole functions.  In the absence of the first-principles theory  
for DBF eqs, it would be quite difficult to give an appropriate model on a reliable 
basis.  \\

\subsection{Conventionality vs. Logical Consistency}

DBF eqs have been popularly used in the macroscopic argument of chiral systems, 
especially in the field of metamaterials research.  As long as they are used for 
nonresonant phenomena as a practical tool, there is not much to say against it, 
except for the difficulty in assigning microscopic meaning to the parameter 
$\beta$.  However, the restriction to the nonresonant phenomena does not 
seem to be widely recognized, to the knowledge of the present author.  In fact 
there are examples of its use for resonant phenomena \cite{Luan, Tomita}.  (The 
constitutive equations used in \cite{Tomita} are  not exactly DBF eqs, but 
$\bVec{D} = \epsilon \bVec{E} - i\xi \bVec{H} \ , \ \bVec{B} = \mu \bVec{H} 
+ i\xi \bVec{E}$, different also from ChC eqs.) 

From the qualitative difference of the two dispersion equations (\ref{eqn:disp1}) 
and (\ref{eqn:disp2}) in resonant region, and from the fact that DBF eqs have no 
support by microscopic theory in contrast to ChC eqs, the use of DBF eqs for 
resonant phenomena is risky.  As a conventional approach with a long history, 
DBF eqs might be kept in use further, but the validity limit should be kept in  
mind. However, if we consider that ChC eqs can be handled as easily as DBF eqs, 
and that they are consistent with the microscopically derived macroscopic 
constitutive equationn, it is highly recommended to use ChC eqs.  For problems 
requiring severe distinction, logical consistency should be preferred to 
conventionality. 
 
\section{Conclusion}

The DBF eqs, popularly used as constitutive equations of chiral media, should be regarded as 
a phenomenological theory applicable only in nonresonant region.  In resonant region, 
it would lead to a qualitatively erroneous result.  On the other hand, the ChC eqs, 
consistent with the first-principles microscopic constitutive equations, can be used both for  
resonant and nonresonant problems. \\

\begin{flushleft}
\underline{Acknowledgment}\\
\end{flushleft}
This work is supported by Grant-in-Aid for scientific research on Innovative Areas 
Electromagnetic Metamaterials of MEXT Japan (Grant No. 22109001).

\end{document}